\UseRawInputEncoding
\documentclass[showpacs,twocolumn,aps,prb]{revtex4-1}
\usepackage{graphicx}
\usepackage{dcolumn}
\usepackage{bm}
\usepackage{amsmath}
\usepackage{amssymb}
\usepackage[usenames,dvipsnames]{xcolor}
\usepackage{color}

\begin{document}


\title{Hexagonal Warping Induced Nonlinear Planar Nernst Effect in Nonmagnetic Topological Insulators}
\author{Xiao-Qin Yu$^{1}$}
\email{yuxiaoqin@hnu.edu.cn}

\author{Zhen-Gang Zhu$^{2,3,4}$}
\email{zgzhu@ucas.ac.cn}

\author{Gang Su$^{3,4,5}$}
\email{gsu@ucas.ac.cn}

\affiliation{$^{1}$ School of Physics and Electronics, Hunan University, Changsha 410082, China.\\
$^{2}$ School of Electronic, Electrical and Communication Engineering, University of Chinese Academy of Sciences, Beijing 100049, China. \\
$^{3}$ Kavli Institute of Theoretical Sciences, University of Chinese Academy of Sciences, Beijing 100049, China.\\
$^{4}$ CAS Center for Excellence in Topological Quantum Computation, University of Chinese Academy of Sciences, Beijing 100190, China.\\
$^{5}$ Theoretical Condensed Matter Physics and Computational Materials Physics Laboratory, College of Physical Sciences, University of Chinese Academy of Sciences, Beijing 100049, China.\\
}

\begin{abstract}
We propose theoretically a new effect, i.e. nonlinear planar Nernst effect (NPNE), in nonmagnetic topological insulator (TI) Bi$_2$Te$_3$ in the presence of an in-plane magnetic field. We find that the Nernst current scales quadratically with temperature gradient but linearly with magnetic field and exhibits a cosine dependence of the orientation of the magnetic field with respect to the direction of the temperature gradient. The NPNE has a quantum origin arising from the conversion of a nonlinear transverse spin current to a charge current due to a joint result of hexagonal warping effect, spin-momentum locking, and the time-reversal symmetry breaking induced by the magnetic field. 
\end{abstract}

\maketitle

\section{Introduction}

The three-dimensional (3D) topological insulators (TI) \cite{M.Z.Hasan,Qi1} represent a new class of 3D materials, owning an insulating bulk and conductive surface states. The surface Dirac electrons have their spin locked perpendicularly to their momenta, namely, spin-momentum locking, giving rise to highly efficient spin-to-charge conversion \cite{Shiomi,Deorani,Wang}, and magnetic switching \cite{Han,Wang2,Dc,Pai} and  great potential application in spintronics and quantum computation \cite{M.Z.Hasan}.

Owing to the spin-momentum-locked surface states, a series of novel magneto-transport properties are identified in nonmagnetic TI film or bilayer structures composed of a ferromagnetic layer and a nonmagnetic TI layer, including both novel linear and nonlinear magnetoelectric effects, such as the non-saturating linear magnetoresistance \cite{Wang3}, the anisotropic magnetoresistance \cite{Wang4,Sulaev}, negative longitudinal magnetoresistance\cite{Wiedmann,Taskin}, bilinear magnetoresistance\cite{Pan1,Dyrdal}, unidirectional magnetoresistance \cite{C.O.Avci1,C.O.Avci2,S.Langenfeld,K.Yasuda1,Y.Lv}, planar Hall effect \cite{Taskin2,Wu,Rakhmilevich,Zheng}, and nonlinear planar Hall effect, etc. The nonlinear planar Hall effect has recently been observed in nonmagnetic TI Bi$_2$Se$_3$ \cite{Pan2}, which describes the Hall resistance linear dependence on both the applied electric field and in-plane magnetic field and is shown to originate from concerted actions of spin-momentum locking  and time-reversal symmetry breaking.

Unlike the extensive exploration on the magneoelectric transport in TIs, only few works have recently focused on the magnetothermal transport. Unidirectional Seebeck effect {\cite{Xiao-Qin-1}}, an nonlinear magnetothermal effect, owing to the asymmetry magnon scattering  was discovered in magnetic TIs, which describes the thermoelectric voltage from Seebeck effect depending on the relative orientations of in-plane magnetization with respect to the temperature gradient.

In this paper, we report another type of nonlinear magnetothermal effect: nonlinear planar Nernst effect (NPNE) in a 3D nonmagnetic TI, i.e. Bi$_2$Te$_3$, in which the Nernst current is quadratically proportional to temperature gradient and linearly proportional to the in-plane magnetic field.  NPNE manifests itself when the applied temperature gradient, magnetic field, and the induced transverse voltage are all coplanar, where the conventional Nernst effect vanishes. Unlike the recently reported topological nonlinear anomalous Nernst effect in strained MoS$_2$ \cite {Xiao-Qin-2} and in bilayer WTe$_2$ \cite{C.Zeng} that origins from Berry curvature in the absence of magnetic field, this nonlinear planar Nernst effect in nonmagnetic TIs is found to originate from the generation of a transverse nonlinear spin current [Fig.\ref{IST} (f)] as second-order response to temperature gradient, which can be converted into a transverse nonlinear planar Nernst current [Fig.\ref{IST} (g)] via in-plane magnetic field collinear with a temperature gradient in the presence of hexagonal warping effect of 2D Fermi contour. We believe that the proposed effect is very useful in magnetotransport and spin caloritronics \cite{Bauer,Avery,Huang,Stephen,Xiao-Qin-3,Baltz}, which is an extension and combination of spintronics and the conventional thermoelectrics, investigating the interplay between a temperature gradient, spin and charge degrees of freedom and aiming at increasing the efficiency and versatility of spin-involved thermoelectric devices.

The paper is organized as follows. We derive the formula of the transverse nonlinear spin current $j^{s}_{y}$ driven by a temperature gradient $\nabla_{x}T$ up to the second order based on the Boltzmann theory in Sec. \ref{NSPNCT}. The expression of NPNE for TI is derived and determined in Sec. \ref{S-NPNE}.  The behavior of NPNE is discussed in Sec. \ref{RAD}. Finally, we give a conclusion in Sec. \ref{CON}.

\section{nonlinear spin Nernst current in Topological insulator}
\label{NSPNCT}
With the relaxation time approximation, the Boltzmann equation for the distribution of electrons in the absence of electric field can be written as
\begin{equation}
f-f_{0}=-\tau \frac{\partial f}{\partial r_{a}}\cdot v_{a}.
\label{Bol-eq3}
\end{equation}
where $\tau$ denotes the relaxation time, and $r_{a}$ and $v_{a}$ represent the $a$ component of coordinate position and velocity of electrons, respectively. $f_{0}=1/\left(\exp\left[\frac{\epsilon\left(\mathbf{k}\right)-E_{f}}{k_{B}T}\right]+1\right)$ is the equilibrium Fermi distribution, where $\epsilon\left(\mathbf{k}\right)$ is energy dispersion, $E_{f}$ indicates the Fermi energy and $k_{B}$ represents Boltzmann constant. The nonequilibrium distribution function response to the second order in temperature gradient can be expanded as $f\approx f_{0}+\delta f_{1}+\delta f_{2}$ with the term $\delta f_{n}$ vanishing as $\left(\partial T/\partial \mathbf{r}_{a}\right)^{n}$. After detail derivation (see Appendix {\ref{ndf}}), the formulas of $\delta f_{1}$ and $\delta f_{2}$ can be determined by Eq. (\ref{Appendix-f1}).

\begin{figure}
\centering
\includegraphics[width=1.0\linewidth,clip]{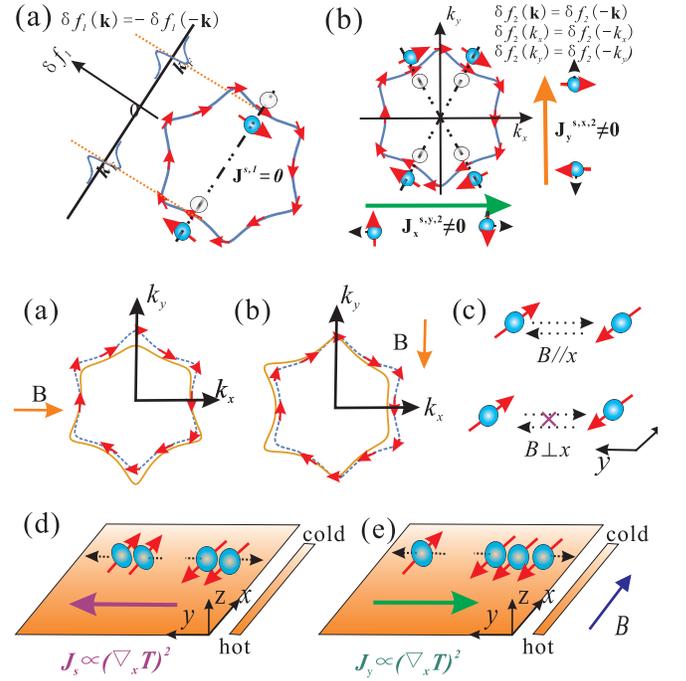}
\caption{(a) The electron distribution along arbitrary $\mathbf{k}$ to the first order of the temperature gradient. (b) Illustration of the generation of a non-equilibrium spin current $j^{s,a_{\bot},2}_{a}$ with spin pointing to $a_{\bot}$-direction to the second order of temperature gradient, where $a=x$ or $y$ and $a_{\bot}$ is orthogonal to $a$ in 2D plane. $\delta f_{1}$ ( $\delta f_{2}$) are the corrections to the equilibrium distribution at the first (second) order of the temperature gradient, respectively. Schematic illustration of the asymmetric distortion of Fermi contour induced by $x$-direction [(c)] and $y$-direction [(d)] magnetic field $B$. The blue dash (yellow solid) curves represent the  Fermi contours of surface band without (with) external magnetic field, respectively. (e) When $B\parallel x$, carriers near the Fermi surface with spin pointing to +$x$ axis or -$x$ axis can be converted into carriers with opposite spin orientation leading to the imbalance of two spin carriers. But when $B\perp x$, the transition between carriers with spin pointing to +$x$ axis or -$x$ axis is forbidden. (f) The generation of the nonlinear spin Nernst current as second order response to the temperature gradient. (g) Schematic illustration for the generation of the nonlinear planar Nernst current when applying an external magnetic field parallel to the temperature gradient.}
\label{IST}
\end{figure}

In the absence of a magnetic field $\mathbf{B}$, the effective Hamiltonian for the surface state of topological insulator \cite{Pan1,Fu,C.Wang} Bi$_{2}$Te$_{3}$ is
\begin{equation}
H_{0}\left(\mathbf{k}\right)=E_{0}\left(k\right)+\boldsymbol{\sigma}\cdot\mathbf{h}\left(\mathbf{k}\right),
\label{H1}
\end{equation}
with
\begin{equation}
\mathbf{h}\left(\mathbf{k}\right)=v_{F}\hbar\mathbf{k}\times\hat{\mathbf{z}}+\lambda\mathbf{k}\times\hat{\mathbf{y}}
\left(k^{2}_{x}-3k^{2}_{y}\right),
\label{h-value1}
\end{equation}
where $\hbar$ is the Plank constant, $v_{F}$ denotes the Fermi velocity, $\boldsymbol{\sigma}$ indicates the Pauli matrices for the two basis functions of the energy bands, and $\lambda$ represents the energy warping parameter. The spin independent term $E_{0}\left(k\right)={\hbar^{2}k^{2}}/{2m^{\ast}}$ generates the particle-hole asymmetry. Unlike the contribution to the nonlinear Hall planar effect\cite{Pan2}, the signal of nonlinear planar Nernst effect arising from the particle-hole asymmetry is insignificant (the details can be found in Appendix \ref{APP-C-PHA}). For simplicity and to emphasize the hexagonal warping effect, we will neglect the particle-hole asymmetry $E_{0}\left(k\right)$ in main text. The second term is the hexagonal warping term which is invariant under threefold rotation $C_{3v}$. $H_{0}\left(\mathbf{k}\right)$ is invariant under the following two operators: 1) mirror reflection $M_{x}$ about the $y$-$z$ plane, and 2) threefold rotation $C_{3}$ about the $z$-axis. The energy eigenvalues
\begin{equation}
\epsilon^{0}_{n}\left(\mathbf{k}\right)=n\sqrt{(v_{F}\hbar k)^{2}+\lambda^{2}k^{6}\cos^{2}3\phi_{k}},
\end{equation}
where $\epsilon^{0}_{n=+1(-1)}$ denotes the energy dispersion of upper (lower) surface bands, respectively, and $\phi_{k}$ is the azimuthal angle of wavevector $\mathbf{k}$ with respect to the $k_{x}$-axis. In the absence of a magnetic field, the time-reversal symmetry is guaranteed, which requires that the energy dispersion respects $\epsilon^{0}_{n}\left({\mathbf{k}}\right)=\epsilon^{0}_n\left({-\mathbf{k}}\right)$ and the mirror symmetry $M_{x}$ imposes the constraint $\epsilon^{0}_{n}\left(k_{x},k_y\right)=\epsilon^{0}_{n}\left(-k_{x},k_y\right)$. Both constraints on the energy dispersion also imply the relation $\epsilon^{0}_{n}\left(k_{x},k_y\right)=\epsilon^{0}_{n}\left(k_{x},-k_y\right)$.  In the following, the upper surface band, namely, $n=1$ will be considered and $\epsilon^{0}_{n=1}\left(\mathbf{k}\right)$ is written as $\epsilon^{0}_{\mathbf{k}}$ for simplicity. The lower surface bands can be analysed in the similar way.

The spin current $j^{s,b}_{a}$ in $a$-direction with spin pointing to the $b$-direction is given by
\begin{equation}
j^{s,b}_{a}=\frac{\hbar}{2}\int[d\mathbf{k}] \langle\sigma^{b}\rangle v_{a}\left(\mathbf{k}\right)f\left(\mathbf{k}\right),
\end{equation}
where $\int [d\mathbf{k}]$ is shorthand for $\int d\mathbf{k}/(2\pi)^{2}$, the average $\langle\cdots\rangle$ is carried out over the surface state of the upper (lower) band and can be replaced by $\langle\sigma^{b}\rangle=n{h^{b}\left(\mathbf{k}\right)}/{h}$ with $\mathbf{h}\left(\mathbf{k}\right)$ defined by Eq. (\ref{h-value1}).

In the absence of a magnetic field, the time reversal symmetry guarantees that the energy dispersion is even in $\mathbf{k}$. i.e., $\epsilon_0\left(\mathbf{k}\right)=\epsilon_0\left(\mathbf{-k}\right)$, which hints that the nonequilibirum electron distribution $\delta f_{1}\sim((\epsilon_\mathbf{k}-E_{f})\partial f_{0}/\partial k_{a})\partial_{a}T$ [Eq. (\ref{Appendix-f1})] in the first order of temperature gradient $\partial_{a}T$ is odd in $\mathbf{k}$, i.e., $\delta f_{1}(-\mathbf{k})= -\delta f_{1}(-\mathbf{k})$, as shown in Fig. \ref{IST}(a). In other words, if the nonequilibrium surface states in $(\mathbf{k},\sigma)$ excesses/deplete due to the first-order variation of temperature gradient, then, the surface states with opposite momentum and spin will deplete/excess, which has no contribution to the spin Nernst current.

\begin{figure}
\centering
 \centering
 \includegraphics[width=0.5\textwidth]{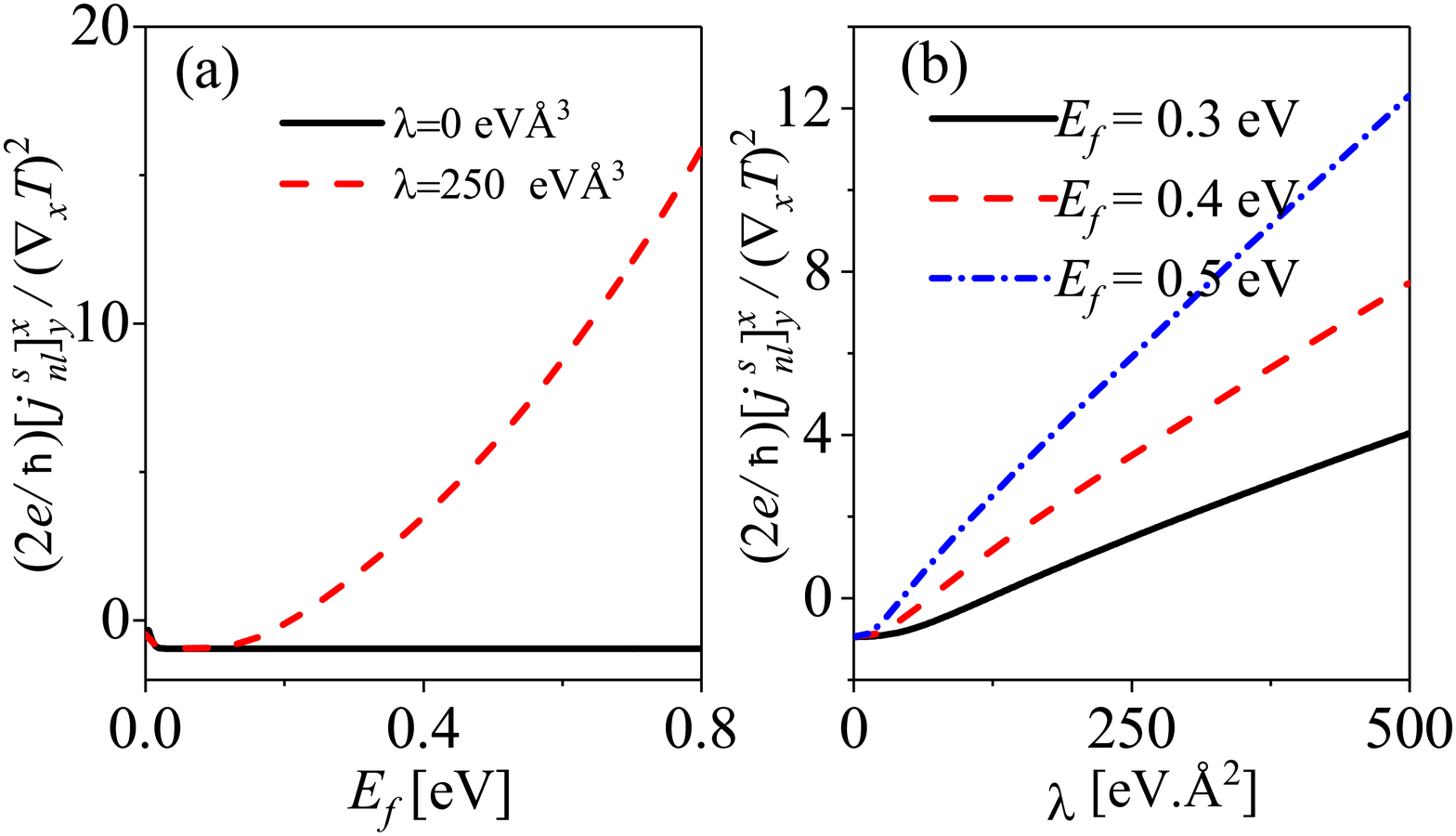}
 \centering
 \includegraphics[width=0.5\textwidth]{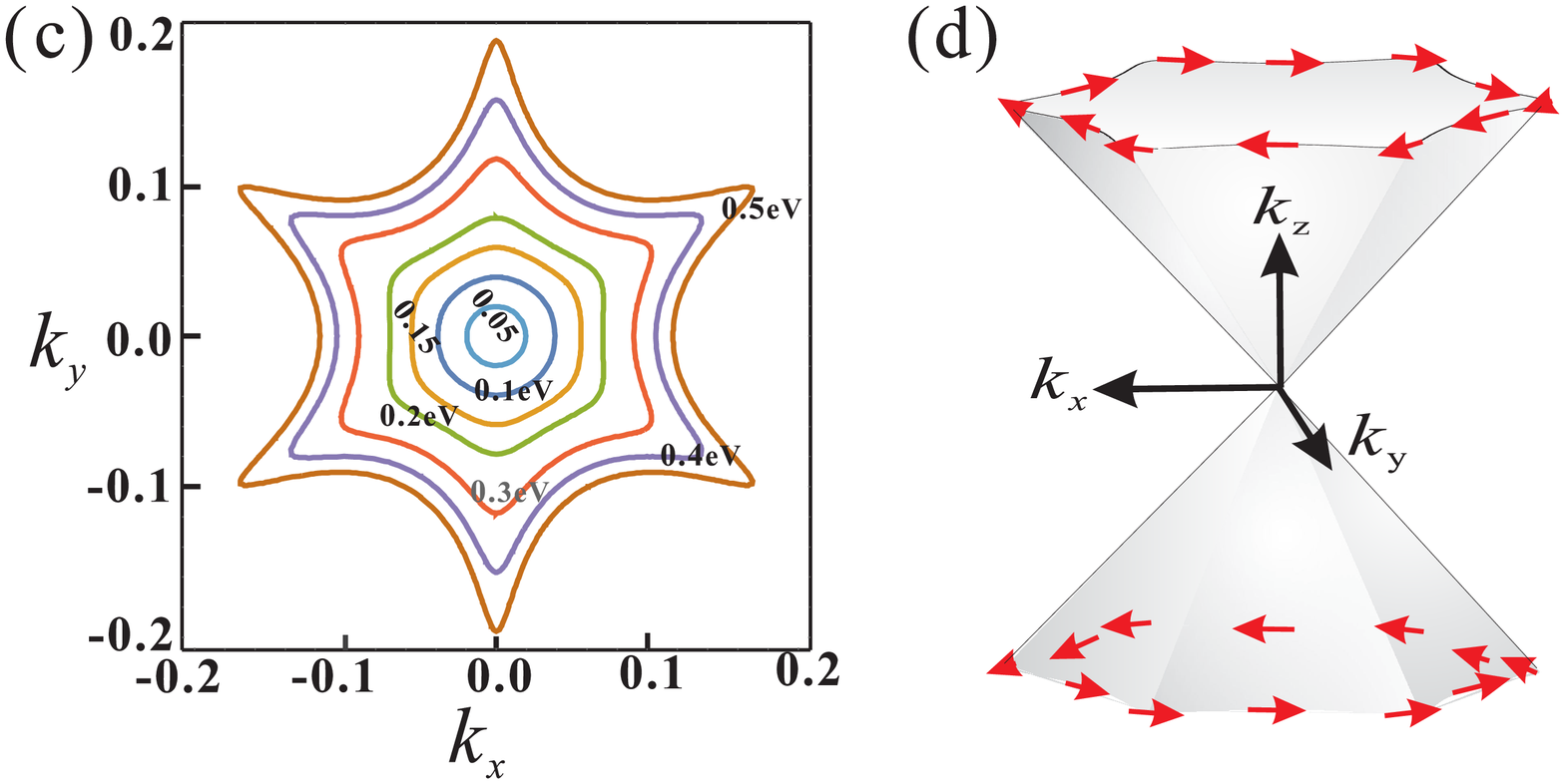}
\caption{The nonlinear spin Nernst current $[{j}^{s}_{nl}]^{x}_{y}$ dependent on Fermi energy $E_{f}$ and hexagonal warping parameter $\lambda$. (a) $\frac{2e}{\hbar}[{j}^{s}_{nl}]^{x}_{y}/(\nabla_{x} T)^{2}$ versus  $E_{f}$ in presence of (in absence of) the hexagonal warping effect. (b) $\frac{2e}{\hbar}[{j}^{s}_{nl}]^{x}_{y}/(\nabla_{x} T)^{2}$ versus $\lambda$ for different Fermi energy. The unit of vertical axis in (a) and (b) is $\text{nA} \mu\text{m}/K^{2}$.
(c) Energy contour of $\epsilon^{0}_\mathbf{k}$ for $\text{Bi}_2\text{Te}_3$. $k_{x}$ and $k_y$ axis are in units of $1 {\AA}$. (d) Schematic depiction of the band structure for the surface states of topological insulator $\text{Bi}_2\text{Te}_3$.}.
\label{F-NSEC}
\end{figure}

On the contrary, the second-order nonequilibrium electron distribution $\delta f_{2}\left(\mathbf{k}\right)$ is even in $\mathbf{k}$. Hence, the nonequilibrium surface states response to the second order of temperature gradient with opposite momentum and opposite spins (due to the spin-momentum locking) are equally populated as shown in Fig. \ref{IST} (b), which leads to a nonzero nonlinear spin current $j^{s,a_{\bot}}_{a}$ with spin orientation in $a_{\bot}$ direction due to the spin-momentum locking, namely, the spins of topological surface states are locked perpendicular to their momenta. Therefore, when applying the temperature gradient in $x$-direction, only nonlinear spin Nernst current $\mathbf{j}^{s}_{nl}$ (where the subscript ``{nl}" and superscript ``{s}" refer to nonlinear and spin, respectively) with spin pointing to $x$-direction gives rise to a transverse spin current in $y$-direction and is found to be
\begin{equation}
\begin{aligned}
\left[\mathbf{j}^{s}_{nl}\right]^{x}_{y}&=\frac{\tau^{2}\hbar}{2}\int[d\mathbf{k}]\frac{(\alpha\hbar k_{y})}{\epsilon^{0}_\mathbf{k}}\left[\frac{\epsilon^{0}_\mathbf{k}-\mu}{T^{2}\hbar}v_{y}v_{x}\frac{\partial f_{0}}{\partial k_{x}}\right.\\
&\left.+\left(\frac{\epsilon^{0}_\mathbf{k}-\mu}{\hbar T}\right)^{2}v_{y}\frac{\partial^{2}f_{0}}{\partial k_{x}^{2} }
\right]\left(\partial_{x}T\right)^{2},\\
\end{aligned}
\end{equation}
where $y$/ $x$ in $\left[\mathbf{j}^{s}_{nl}\right]^{x}_{y}$ indicates the movement direction of carrier/ spin orientation, respectively. This nonlinear spin Nernst current originated from the topological surface states could be a source of spin injection and spin current generation in future applications of spin caloritronics.

A set of constant energy contours of $H_{0}\left(\mathbf{k}\right)$ are be obtained, as plotted in Fig. \ref{F-NSEC}(c), where we have taken $\lambda=250\, \text{eV} {\AA}^{3}$ and $v_{F}\hbar=2.25\,{\color{blue}{\text{eV} {\AA}}}$ for Bi$_2$Te$_3$ \cite{Fu}. When the Fermi energy gets close to the Dirac point ($E=0$  eV), the Fermi surface manifests itself as a circle and the warping effect is inapparent. The Fermi surface starts to deviate considerably from a circle and becomes more hexagonalike around $E=0.2$ eV.

Figs. \ref{F-NSEC}(a) and (b) illustrate the dependence of nonlinear spin Nernst current (NSNC) $\left[\mathbf{j}^{s}_{nl}\right]^{x}_{y}$  on the Fermi energy and the hexagonal warping effect. A larger $\left[\mathbf{j}^{s}_{nl}\right]^{x}_{y}$ can be generated by increasing the hexagonal warping parameters and the absolute value of the Fermi energy in which the hexagonal warping effect will be enhanced. An interesting finding is that in addition to the contribution of hexagonal warping term, the linear-$k$ Dirac dispersion ($\lambda=0$) can also give rise to the signal of nonlinear spin Nernst current, which is distinguished from the electric-field-induced nonlinear spin Hall current\cite{Pan2}. This can explain why the NSNC is nonzero when the energy is in the range of [0 eV, 0.2eV] [Fig. \ref{F-NSEC}(a)], in which the trigonal warping effect is insignificant [Fig. \ref{F-NSEC}(c)].  However, the signal of NSNC originated from the linear-$k$ Dirac dispersion cannot be converted into the nonlinear planar Nernst current when the Fermi energy is away from the Dirac point (see the details in Sec. \ref{S-NPNE} and Appendix \ref{CLDN}).

\section{nonlinear planar Nernst effect in topological insulator}
\label{S-NPNE}
In the absence of magnetic field, the carriers with opposite spins are equally populated and move in opposite directions in transverse direction ($y$-direction) [Fig. \ref{IST}(f)]. Hence, there is no charge current flux vertical to temperature gradient. However, when applying an in-plane magnetic field to the topological insulator, because of the spin-momentum locking, the Fermi surface will be distorted in the direction perpendicular to magnetic field [Figs. \ref{IST}(c) and (d)] due to the hexagonal warping term, which leads to the imbalance between the two spin fluxes of the spin current and thus, the spin current is partially converted into the nonlinear planar Nernst current (NPNC) [Fig. \ref{IST}(g)].

It should be emphasized that the successful conversion from the spin current into NPNC is ensured by the hexagonal warping effect. If there is no hexagonal warping term, i.e., $\lambda=0$, the energy dispersion turns into the linear-$k$ Dirac dispersion and the Fermi surface returns to a circle. Instead of being distorted, pervious studies \cite{K.Yasuda1,Xiao-Qin-1} show that the whole linear dispersion will shift in the momentum space when applying an in-plane magnetic field. Thus, the spin population will stay the same and the two spin fluxes still keep the balance, hinting that there is no NPNC in this case. However, one might notice that there is very weak signal (almost $500$ times smaller than the sign from the warping effect, see Fig.\ref{Q-FT}(a)) stemmed from the linear dispersion when the Fermi energy is located near  Dirac point within $10 k_{B}T$. This weak signal can be attributed to the temperature broadening effect (see Appendix \ref{CLDN} for a detailed discussion).

In the presence of a magnetic field $\mathbf{B}$, the effective Hamiltonian for the surface state of topological insulator Bi$_{2}$Te$_{3}$ \cite{Fu} is given by
\begin{equation}
\begin{aligned}
H\left(\mathbf{k}\right)=\boldsymbol{\sigma}\cdot\left[\mathbf{h}\left(\mathbf{k}\right)+g\mu_{B}\mathbf{B}\right],
\end{aligned}
\label{H1}
\end{equation}
where $g$ and $\mu_{B}$ represent the $g$-factor and Bohr magneton, respectively. The energy eigenvalues are
\begin{equation}
\epsilon^\text{M}_n\left(\mathbf{k}\right)=n|\mathbf{h}\left(\mathbf{k}\right)+g \mu_{B}\mathbf{B}|.
\label{ED}
\end{equation}
In the following, we shall consider the upper surface bands, namely $n=1$, and write $\epsilon^\text{M}_{n=1}\left(\mathbf{k}\right)$ as $\epsilon^\text{M}_{\mathbf{k}}$ for simplicity. The lower surface bands can be analysed in the similar way.

\begin{table}[tbph]
\centering
\caption{ Parity about $k_{x}$ or $k_{y}$ for Dirac dispersion of topological insulator in the absence of a magnetic field.}
\begin{tabular*}{8 cm}{@{\extracolsep{\fill}}lccccc}
 \hline\hline
 \text{function} &\text{parity for} $k_x$ &\text{parity for} $k_y$\\
 $\epsilon^{0}_{\mathbf{k}}$ &\text{even} &\text{even}\\
 $v_{x}$ &\text{odd} &\text{even}\\
 $v_{y}$ &\text{even} &\text{odd}\\
 $\frac{\partial f_{0}}{\partial k_{x}}$ &\text{odd} &\text{even}\\
 $\frac{\partial f_{0}}{\partial k_{y}}$ &\text{even} &\text{odd}\\
 \hline\hline
 \end{tabular*}
 \label{parities}
 \end{table}

The charge current $j_{a}$ in $a$-direction is $j_{a}=-e\int[d\mathbf{k}]v_{a}f(\mathbf{r},\mathbf{k})$.  After tedious derivation in Appendix {\ref{npl}}, the current $j_{a}^{\left(1\right)}$ and $j_{a}^{\left(2\right)}$ as the first-order and second-order responses to the temperature gradient in the first-order approximation of magnetic field are found, respectively, to be

\begin{equation}
\begin{aligned}
j^{\left(1\right)}_{a}&=\sum_{b}G_{ab}\partial_{b}T+\sum_{bc}K_{abc}\partial_{b}TB_{c},\\
j^{\left(2\right)}_{a}&=\sum_{bc}W_{abc}\partial_{b}T\partial_{c}T+\sum_{bcd}Q_{abcd}\partial_{b}T\partial_{c}TB_{d},
\end{aligned}
\end{equation}
where the relation $\partial\epsilon_\mathbf{k}^\text{M}/\partial B_{d}=g\mu_{B}\partial\epsilon_\mathbf{k}^\text{M}/\partial h_{d}$ has been applied. Explicit expressions for the linear current response ($G_{ab}$, $K_{abc}$) and nonlinear response function ($W_{abc}$, $Q_{abcd}$) are given in Eqs. (\ref{App-B-cu-f1}) and (\ref{App-B-cu-f2}).

Through exploiting the parity in Table \ref{parities}, one can find the following tensor elements are zero, i.e.,
\begin{equation}
\begin{aligned}
G_{xy}&=G_{yx}=0,\\
K_{abc}&=0,\, W_{abc}=0, \quad a,b,c=x,y\\
Q_{xyyx}&=Q_{xyxy}=Q_{xxyy}=Q_{xxxy}=0,\\
Q_{yxxy}&=Q_{yxyx}=Q_{yyxx}=0,\\
\end{aligned}
\label{coefficients}
\end{equation}
which suggest that when applying an in-plane magnetic field  $\mathbf{B}=B(\cos \theta,\sin\theta)$ and temperature gradient $\partial_{x} T$ along $x$-direction (i.e., $b=c=x$), the planar Nernst effect $j^{\left(1\right)}_{a}$ disappears in Bi$_2$Te$_3$  and has no contribution to the transverse thermal voltage signal. 
And the current density $j^{\left(2\right)}_{y}$ flowing along the $y$-direction (i.e., $d=y$) as the response to the second order in temperature gradient stems from the nonlinear planar Nernst current density $j^\text{p}_{nl}$ (where the subscript ``{nl}" and superscript ``{p}" denote nonlinear and planar, respectively) and is found to be
\begin{equation}
\begin{aligned}
j^{\left(2\right)}_{y}&=j^\text{p}_{nl}=\left(Q_{yxxx}\cos\theta+Q_{yxxy}\sin\theta\right)
\left(\partial_{x}T\right)^{2}B\\
&=Q_{yxxx}\cos \theta \left(\partial_{x}T\right)^{2}B,
\end{aligned}
\label{current1}
\end{equation}
where the nonlinear planar coefficient $Q_{yxxx}$ is given as
\begin{equation}
\begin{aligned}
Q_{yxxx}&=-\frac{e\tau^{2}gu_{B}}{\alpha T^{2}\hbar^{2}}\int\left[\mathbf{dk}\right]\left[\frac{\partial f_{0}}{\partial \epsilon^{0}_\mathbf{k}}\hbar^{2}\Upsilon_{1}+\left(\epsilon^{0}_{k}-\mu\right)\left(\frac{\partial f_{0}}{\partial \epsilon^{0}_\mathbf{k}}\right.\right.\\
&\left.\times\hbar\Upsilon_{2}+3\frac{\partial^{2} f_{0}}{\partial \left(\epsilon^{0}_k\right)^{2}}\hbar^{2}\Upsilon_{1}\right)
+\left(\epsilon^{0}_\mathbf{k}-\mu\right)^{2}\left(\frac{\partial f_{0}}{\partial \epsilon^{0}_\mathbf{k}}\Upsilon_{3}\right.\\
&\left.\left.+\frac{\partial^{2} f_{0}}{\partial \left(\epsilon^{0}_\mathbf{k}\right)^{2}}\hbar\Upsilon_{4}+\frac{\partial^{3} f_{0}}{\partial \left(\epsilon^{0}_\mathbf{k}\right)^{3}}\hbar^{2}\Upsilon_{1}\right)\right],
\end{aligned}
\label{Q_yxxx}l
\end{equation}
where the coefficients $\Upsilon_{1}$,$\Upsilon_{2}$,$\Upsilon_{3}$ and $\Upsilon_{4}$ are given in {\color{blue}{ Eq. (\ref{App-B-Ga})}}.

\section{Results and Discussion}
\label{RAD}
Eq. (\ref{current1}) indicates the nonlinear planar current $j^\text{p}_{nl}$ exhibits $\cos\theta$ dependence on the orientation of magnetic field and is proportional to the $x$-component of the magnetic field $B_{x}\propto B\cos\theta$. Thus, when the magnetic field is collinear with the temperature gradient (i.e., $\theta=0,\pi,2\pi$), the magnitude of $\mid j^\text{p}_{nl}\mid$ will reach its maximum. However, the nonlinear planar Nernst effect will disappear when the magnetic field $\mathbf{B}$ is vertical to the temperature gradient. These features of the nonlinear planar Nernst current depending on the orientation of magnetic field can be ascribed to the spin-momentum locking. As shown in Fig. \ref{IST}(f) and discussion in Sec. \ref{NSPNCT}, the spin orientation in the nonlinear spin Nernst current generated by temperature gradient $\nabla_{x} T$ is along $x$-direction. Therefore, only the $x$-component of magnetic field can lead to a transition between the two spin currents and induces the imbalance of two spin carriers [Fig. \ref{IST}(e)]. As a result the nonlinear spin Nernst current will be partially converted to nonlinear planar Nernst current [Figs. \ref{IST}(f) and (g)].

\begin{figure}
\centering
\includegraphics[width=1.0\linewidth,clip]{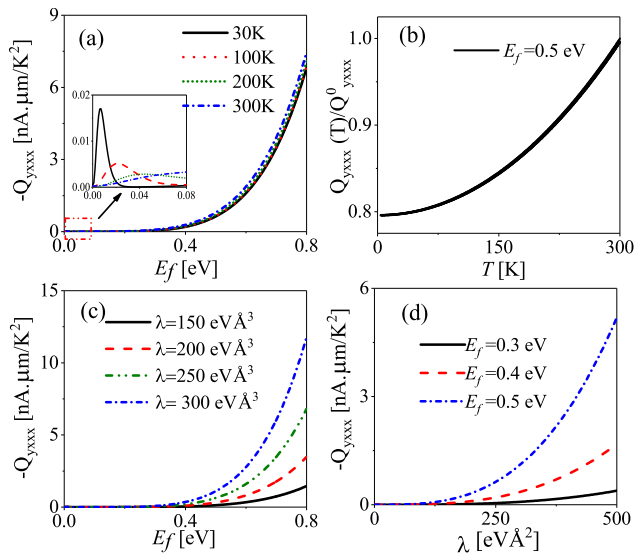}
\caption{The nonlinear planar coefficient (NPC) $Q_{yxxx}$ [(a) and (c)] as a function of Fermi energy for different temperature and different energy warping parameter $\lambda$. (b) $Q_{yxxx}(T)/Q_{yxxx}^{0}$ versus temperature. $Q_{yxxx}^{0}$ is the NPC for $T=300\text{K}$ . (d) $Q_{yxxx}$ as a function of $\lambda$ at different Fermi energy. The energy warping parameter $\lambda$ is taken 250 $\text{eV} {\AA}^{3}$ in (a) and (b). $T=30 \text{K}$ is fixed in (c) and (d). Parameters used: $v_{F}\hbar=2.25 \text{eV}{\AA}$, $g=2$  and $\tau=5.864\times10^{-13}s $. Here, all parameters are taken from topological insulator Bi$_2$Te$_3$. }
\label{Q-FT}
\end{figure}

We use the following parameters for Bi$_{2}$Te$_{3}$: the Fermi velocity $v_{F}\hbar=2.25\,\text{eV} {\AA}$, $g=2$, and the scattering relaxation time $\tau\approx 5.864\times 10^{-13}s$  is estimated by $\tau=\mu m/e$. The mobility of surface states in Bi$_{2}$Te$_{3}$ can range from $9\times 10^{3}$ to $10^{4}$ cm$^{2}$V$^{-1}$s$^{-1}$.\cite{Qu} $\mu=9000$ cm$^{2}$V$^{-1}$s$^{-1}$ is used for an estimation.


It's observed that a very weak signal appears near the Dirac point with a few $k_{B}T$, and $Q_{yxxx}$ is almost zero when the energy is in the range of [0, 0.2eV], as expected, since the trigonal warping effect is insignificant and the Fermi surface almost displays like a circle [Fig. \ref{F-NSEC}(c)] in this range. The appearance of faint signal at the Dirac point can attributes to the thermal broadening effect of nonequilibrium Fermi distribution near Dirac point for the linear-$k$ Dirac dispersion [the details can be found in Appendix \ref{CLDN}]. Besides, one might notice that the signal of $Q_{yxxx}$ is still quite weak when the Fermi energy is in the range of [0.2, 0.4] eV, a regime where a warped Fermi surface is present [Fig. \ref{F-NSEC}] (c). This can be attributed to the low conversion efficiency from the nonlinear spin to charge current [Fig. \ref{CR}]. However, when the absolute value of Fermi energy $|E_{f}|$ is increased sufficiently, the trigonal warping effect will become profound and lead to a large enhancement of nonlinear planar Nernst effect. It is interesting to point out that the impact of varying temperature is negligible [Fig. \ref{Q-FT}(a) (b)] when Fermi energy is away from Dirac point. 
Figure \ref{Q-FT}(c) and $(d)$ present the Fermi energy and hexagonal warping dependence of  $Q_{yxxx}$. The magnitude of $Q_{yxxx}$ increases monotonously with the enhanced energy warping parameter $\lambda$. As expected, when $\lambda$ tends to be zero, the nonlinear planar Nernst effect will disappear.

To numerically estimate the proposed effect, we take $Q_{yxxx}\approx0.8 \,\text{nA}\cdot \mu\text {m}/\text{T K}^{2}$ [Fig. \ref{Q-FT}(b)] for $T=30 \text{K}$ and $E_{f}=0.5\, eV$. In experiment, the temperature gradient can already reach $1 \,\text{K}\mu\text{m}^{-1}$.\cite{Xu} Therefore, when applying the magnetic field $B=3\text{T}$ parallel to temperature gradient, the nonlinear planar Nernst current $j^\text{P}_{nl}\times l$ [Eq. (\ref{current1})] of Bi$_2$Te$_3$ is estimated to be order of  $0.16\, \mu\text{A}$ with the length of sample $l=50\,\mu m$, which is measurable.

A Rashba-split surface states in two-dimensional electron gas (2DEG) \cite{Y.Wang,Mellnik,Denkert} [Fig. \ref{App-c3}(d)] might coexists with topological surfaces states (TSS) due to the surface band bending in topological insulators, which may also have a significant contribution to the nonlinear Planar Nernst effect. However, it is found that only when Fermi energy locates near the Lifshitz point within a few $k_{B}T$, a very weak signal ($100$ times smaller than the signal arising from TSS) can be generated [see the details in Appendix \ref{COR2D}]. Therefore, the contribution of Rashba 2DEG to NPNE can be neglected.

\section{Conclusion}
\label{CON}
In summary, we propose a new effect, i.e. the nonlinear planar Nernst effect (NPNE) in this work. It is found that a nonlinear spin-Nernst current, originated from the hexagonal warping effect and the nonequilibrium carrier distribution, flows transversely to temperature gradient direction and can be partially converted into the nonlinear-planar-Nernst current $j^\text{p}_{nl}$ when an in-plane magnetic field is applied to TI.  The quantity of $j^\text{p}_{nl}$ is strongly dependent on the orientation of the magnetic field. When the in-plane magnetic field is collinear to the temperature gradient, $|j^\text{p}_{nl}|$ will reach its maximum. However, $j^\text{p}_{nl}$ becomes zero when the magnetic field is perpendicular to the temperature gradient. The magnitude of NPNE is strongly affected by the hexagonal warping term and the Fermi energy. Except a very faint signal of NPNE appearing near the Dirac point within a few $k_{B}T$ due to the temperature broadening effect, when the Fermi level is close to the Dirac point, the signal of the NPNE mostly disappears due to the weak hexagonal warping effect.  However, when enlarging the value $|E_{f}|$ sufficiently, the NPNE rapidly increases owing to the profound hexagonal warping effect. 
 The nonlinear planar Nernst effect proposed here might also occur in other noncentrosymmetric materials with strong spin-orbit coupling and nontrivial spin textures. Therefore, our findings have great potential application in magneto-thermal transport and spin caloritronics, and might pave a new way to the emerging field of nonlinear spin caloritronics.
\\

This work is supported by the Fundamental Research Funds for the Central Universities and the NSFC (Grant No.12004107). G.S. and Z.G.Z. are supported in part by the National Key R\&D Program of China (Grant No. 2018FYA0305800), the Strategic Priority Research Program of CAS (Grant Nos. XDB28000000),  the NSFC (Grant No. 11834014), and Beijing Municipal Science and Technology Commission (Grant No. Z118100004218001). Z.G.Z. is also supported in part by the NSFC (Grant Nos. 11674317 and 11974348).

\appendix
\setcounter{equation}{0}
\setcounter{figure}{0}
\setcounter{table}{0}
\makeatletter
\renewcommand{\theequation}{A\arabic{equation}}
\renewcommand{\thefigure}{A\arabic{figure}}
\renewcommand{\thetable}{A\arabic{table}}

\bigskip
\bigskip

\noindent
\section{The non-equilibrium distribution function in the presence of temperature gradient} \label{ndf}

With the relaxation time approximation, the Boltzmann equation for the distribution of electrons in the absence of an electric field can be written as
\begin{equation}
\frac{\partial f}{\partial r_{a}}\cdot v_{a}+\frac{e}{\hbar}(\vec{v}\times\vec{B})\cdot \frac{\partial {f}}{\partial \vec{k}}=-\frac{f-f_{0}}{\tau}.
\label{App-Bol-eq1}
\end{equation}
In two-dimensional (2D) transport, the Lorentz force has no contribution to the electron dynamics for the in-plane magnetic field because of $(\vec{v}\times\vec{B})\cdot \frac{\partial {f}}{\partial \vec{k}}=0$. Thus, in the presence of an in-plane magnetic field the Boltzmann equation in Eq. (\ref{App-Bol-eq1}) for 2D transport can be further simplified as
\begin{equation}
f-f_{0}=-\tau \frac{\partial f}{\partial r_{a}}\cdot v_{a}.
\label{Bol-eq3}
\end{equation}
To the response up to the second order in temperature gradient $\nabla T$, the local distribution function $f\left(\mathbf{r},\mathbf{k}\right)$ can be expanded as
\begin{equation}
\begin{aligned}
f\left(\mathbf{k},\mathbf{r}\right)=&f_{0}\left(\mathbf{k},\mathbf{r}\right)+A_{a}\frac{\partial T}{\partial r_{a}}+
B_{a \beta}\frac{\partial{T}}{\partial{r_{a}}}\frac{\partial{T}}{\partial{r_{b}}}+O[3]\\
\approx&f_{0}\left(\mathbf{k},\mathbf{r}\right)+\delta f_{1}\left(\partial_{a}T\right)+\delta f_{2}\left(\partial_{a}T\partial_{b}T\right),\\
\end{aligned}
\label{non-equ-f-1}
\end{equation}
with
\begin{equation}
\left\{
\begin{aligned}
\delta f_{1}\left(\partial_{a}T\right)&=A_{a}\partial_{a}T,\\
\delta f_{2}\left(\partial_{a}T\partial_{b}T\right)&=B_{a b}\partial_{a}T\partial_{b}T,\\
\partial_{a}&\rightarrow \frac{\partial}{\partial r_{a}}.
\end{aligned}
\right.
\end{equation}
where $f_{0}\left(\mathbf{k},\mathbf{r}\right)$ is the local equilibrium distribution, which is itself fixed by the temperature at $\mathbf{r}$ \cite{Ziman}, giving
\begin{equation}
\frac{\partial f_{0}}{\partial r_{a}}=\frac{\partial f_{0}}{\partial T}\frac{\partial T}{\partial r_{a}}=-\frac{\left(\epsilon_\mathbf{k}-\mu_e\right)}{T}\frac{\partial f_{0}}{\partial \epsilon_\mathbf{k}}\frac{\partial T}{\partial r_{a}}.
\end{equation}
 Substituting the formula of $f$ in Eq. (\ref{non-equ-f-1}) into Eq. (\ref{Bol-eq3}) and comparing the expansion coefficients in the first-order of $\partial_{a} T$, one obtains
\begin{equation}
\begin{aligned}
\delta f_{1}\left(\partial_{a}T\right)
=&-\tau\frac{\partial f_{0}}{\partial r_{a}}\cdot v_{a}+O[\partial_{a} T\partial_{b}T].\\
\end{aligned}
\end{equation}
Thus, we can have
\begin{equation}
\delta f_{1}\left(\partial_{a}T\right)=-\tau\frac{\partial f_{0}}{\partial T}\partial_{a}T\cdot v_{a}.
\label{non-equ-f1}
\end{equation}
By iteration, then, we can have
\begin{equation}
\begin{aligned}
&\delta f_{2}\left(\partial_{a}T\partial_{b}T\right)\\
=&-\tau\frac{\partial \delta f_{1}}{\partial r_{a}}\cdot v_{a}\\
=&{\tau^{2}}\left(\frac{\partial^{2}f_{0}}
{\partial^{2}T}{\partial_{a}T}{\partial_{b}T}+\frac{\partial f_{0}}{\partial T}\partial_{ab}T\right)v_{b}v_{a}.
\end{aligned}
\label{f22}
\end{equation}

Here, we introduce a trick to transform $\frac{\partial f_{0}}{\partial T}$ into $\frac{\partial f_{0}}{\partial \mathbf{k}}$ 
through a partial differential treatment,
\begin{equation}
\begin{aligned}
\frac{\partial f_{0}}{\partial \mathbf{k}}=\frac{\partial f_{0}}{\partial \epsilon_\mathbf{k}}\cdot \frac{\partial \epsilon_\mathbf{k}}{\partial \mathbf{k}}
=-\frac{\partial f_{0}}{\partial T}\frac{\hbar\mathbf{v}T}{\left(\epsilon_\mathbf{k}-\mu_e\right)}.
\end{aligned}
\label{IK-tran}
\end{equation}

In the above, we have used the relation: $\frac{\partial f_{0}}{\partial T}=-\frac{\left(\epsilon_\mathbf{k}-\mu_e\right)}{T}\frac{\partial f_{0}}{\partial \epsilon_\mathbf{k}}$ and $\frac{\partial \epsilon_\mathbf{k}}{\partial \mathbf{k}}={\hbar \mathbf{v}}$.

From Eq. (\ref{IK-tran}), it is easily to obtain the following identities:
\begin{equation}
\begin{aligned}
\frac{\partial{f_{0}}}{\partial T} \cdot v_{a}&\!=\!-\frac{\epsilon_\mathbf{k}-\mu_e}{\hbar T}\frac{\partial f_{0}}{\partial k_{a}},\\
\frac{\partial^{2}f_{0}}{\partial T^{2}}v_{a}v_{b}&\!=\!\frac{E_{k}\!-\!\mu_e}{\hbar T^{2}}\frac{\partial f_{0}}{\partial k_{a}}v_{b}\!+\!\left(\frac{E_{k}\!-\!\mu_e}{\hbar T}\right)^{2}\frac{\partial^{2} f_{0}}{\partial k_{a}\partial k_{b}}.
\end{aligned}
\end{equation}
Taking these identities into the formulas of $\delta f_{1}$ [Eq. (\ref{non-equ-f1})] and $\delta f_{2}$ [Eq. (\ref{f22})] and assuming the uniform temperature gradient in the system, i.e., $\partial_{ab}T=0$, one obtains
\begin{equation}
\begin{aligned}
\delta f_{1}&=\frac{\tau}{T\hbar}\left(\epsilon_\mathbf{k}-\mu_e\right)\frac{\partial f_{0}}{\partial k_{a}} \partial_{a} T,\\
\delta f_{2}&=\frac{\tau^{2}}{T^{2}\hbar^{2}}\left(\hbar v_{b}\frac{\partial f_{0}}{\partial k_{a}}+\left(\epsilon_{k}-\mu_e\right)\frac{\partial^{2} f_{0}}{\partial k_{a}\partial k_{b}}\right)\\
&\times\left(\epsilon_\mathbf{k}-\mu_e\right)\partial_{a}T\partial_{b}T.
\end{aligned}
\label{Appendix-f1}
\end{equation}

\makeatletter
\renewcommand{\theequation}{B\arabic{equation}}
\renewcommand{\thefigure}{B\arabic{figure}}
\renewcommand{\thetable}{B\arabic{table}}
\section{The formula of nonlinear planar current for topological insulator} \label{npl}
Based on Eq. (\ref{Appendix-f1}), one can determine the charge current $j_{a}=-e\int[d\mathbf{k}]v_{a}f(\mathbf{r},\mathbf{k})$ in $a$-direction as the first-order and second-order responses to the temperature gradient, respectively, as
\begin{equation}
\begin{aligned}
j_{a}^{\left(1\right)}&=-\tau e\int[d\mathbf{k}]\frac{\epsilon_{k}-\mu}{T\hbar}v_{a}\frac{\partial f_{0}}{\partial k_{b}}
\partial_{b}T,\\
j_{a}^{\left(2\right)}&=-\tau^{2}e\int[d\mathbf{k}]\left[\frac{\epsilon_{k}-\mu}{T^{2}\hbar}v_{a}v_{b}\frac{\partial f_{0}}{\partial k_{c}}\right.\\
&\left.+\left(\frac{\epsilon_{k}-\mu}{\hbar T}\right)^{2}v_{a}\frac{\partial^{2}f_{0}}{\partial k_{b}\partial k_{c}}
\right]\partial_{b}T\partial_{c}T.\\
\end{aligned}
\label{cu-f1}
\end{equation}

In presence of a magnetic field, the energy dispersion $\epsilon^\text{M}_{n}\left(\mathbf{k}\right)$ for nonmagnetic topological insulator Bi$_2$Te$_3$  is given in Eq.(\ref{ED}). We only consider the upper surface band, and write $\epsilon^\text{M}_{n}\left(\mathbf{k}\right)$ as $\epsilon^\text{M}_\mathbf{k}$. One can find $
{\partial\epsilon^\text{M}_\mathbf{k}}/{\partial B_{d}}=gu_{B}{\partial\epsilon^\text{M}_\mathbf{k}}/{\partial h_{d}}$,
which hints
\begin{equation}
\frac{\partial F\left(\epsilon^\text{M}_\mathbf{k}\right)}{\partial B_{d}}=gu_{B}\frac{\partial F \left(\epsilon^\text{M}_\mathbf{k}\right)}{\partial h_{d}}.
\label{Pro}
\end{equation}

Therefore, to the first order of magnetic field, the current $j_{a}^{\left(1\right)}$ and $j_{a}^{\left(2\right)}$ in Eq. (\ref{cu-f1}) is found to be
\begin{equation}
\begin{aligned}
j^{\left(1\right)}_{a}&=\sum_{b}G_{ab}\partial_{b}T+\sum_{bc}K_{abc}\partial_{b}TB_{c},\\
j^{\left(2\right)}_{a}&=\sum_{bc}W_{abc}\partial_{b}T\partial_{c}T+\sum_{bcd}Q_{abcd}\partial_{b}T\partial_{c}TB_{d},
\end{aligned}
\end{equation}
with
\begin{eqnarray}
G_{ab}&=&-\frac{\tau e}{\hbar T}\int[d\mathbf{k}]\left(\epsilon^{0}_\mathbf{k}-\mu\right)v_{a}\frac{\partial f_{0}}{\partial k_{b}},\nonumber\\
K_{abc}&=&-\frac{\tau e gu_{B}}{\hbar T}\int[d\mathbf{k}] \left[v_{a}\frac{\partial \epsilon^{0}_{k}}{\partial h_{c}}\frac{\partial f_{0}}{\partial k_{b}}+\left(\epsilon^{0}_{k}-\mu\right)\right.\nonumber\\
&&\left.\times \left(\frac{\partial v_{a}}{\partial h_{c}}\frac{\partial f_{0}}{\partial k_{b}}+v_{a}\frac{\partial^{2} f_{0}}{\partial h_{c}\partial k_{b}}\right)\right],\nonumber\\
W_{abc}&=&-\frac{\tau^{2}e}{T^{2}\hbar^{2}}\int[d\mathbf{k}]v_{a}\left[\left(\epsilon^{0}_{k}-\mu\right)\hbar v_{b}\frac{\partial f_{0}}{\partial k_{c}}\right.\nonumber\\
&&\left.+\left({\epsilon^{0}_{k}-\mu}\right)^{2}\frac{\partial^{2}f_{0}}{\partial k_{b}\partial k_{c}}
\right],\nonumber\\
\label{App-B-cu-f1}
\end{eqnarray}
\begin{widetext}
\begin{equation}
\begin{aligned}
Q_{abcd}=&\frac{-\tau^{2}egu_{B}}{T^{2}\hbar^{2}}\int[d\mathbf{k}]\left[\left({\epsilon^{0}_{k}-\mu}\right)^{2}\left(\frac{\partial^{2} f_{0}}{\partial k_{b}\partial k_{c}}\frac{\partial v_{a}}{\partial h_{d}}+\frac{v_{a}\partial^{3} f_{0}}{\partial k_{b}\partial k_{c}\partial h_{d}}\right)+\frac{\partial \epsilon^{0}_{k}}{\partial h_{d}}\hbar v_{a}v_{b}\frac{\partial f_{0}}{\partial k_{c}}+2\left(\epsilon^{0}_{k}-\mu\right)\frac{\partial \epsilon^{0}_{k}}{\partial h_{d}}v_{a}\frac{\partial^{2}f_{0}}{\partial k_{b}\partial k_{c}}\right.\\
&\left.+\left({\epsilon^{0}_{k}-\mu}\right)\hbar\left(\frac{\partial^{2} f_{0}}{\partial k_{c}\partial h_{d}}v_{a}v_{b}+\frac{\partial f_{0}}{\partial k_{c}}\frac{\partial v_{a}}{\partial h_{d}}v_{b}+\frac{\partial f_{0}}{\partial k_{c}}\frac{\partial v_{b}}{\partial h_{d}}v_{a}\right)\right].
\end{aligned}
\label{App-B-cu-f2}
\end{equation}
\end{widetext}
To obtain Eqs.(\ref{App-B-cu-f1}) and (\ref{App-B-cu-f2}), we have used the relation $\frac{\partial F\left(\epsilon^\text{M}_\mathbf{k}\right)}{\partial B_{d}}=gu_{B}\frac{\partial F \left(\epsilon^\text{M}_\mathbf{k}\right)}{\partial h_{d}}$ with $h_{d} \left(d=x,\,y\, \text{or} z\right)$, $\epsilon^{0}_\mathbf{k}=|\mathbf{h}\left(\mathbf{k}\right)|$ is the eigenvalues for the effective Hamiltonian $H^{\left(0\right)}=\boldsymbol{\sigma}\cdot\mathbf{h}\left(\mathbf{k}\right)$. According to the formulas $\mathbf{h}\left(\mathbf{k}\right)$ in Eq. (\ref{h-value1}), one can obtain

\begin{equation}
\frac{\partial}{\partial h_{x}}=\frac{\partial}{(\alpha \hbar\partial k_{y})}, \quad \frac{\partial}{\partial h_{y}}=\frac{\partial}{(\alpha \hbar\partial k_{x})}.
\label{App-B-Rel}
\end{equation}

When applying an in-plane magnetic field $\mathbf{B}=B(\cos \theta,\sin\theta)$ and temperature gradient $\partial_{x} T$ along $x$-direction (i.e., $b=c=x$), the planar Nernst current density $j^{\left(1\right)}_{y}$ and nonlinear planar Nernst current density $j^{\left(2\right)}_{y}$ in y-direction (i.e., $d=y$), as the response to the first order and the second order in temperature gradient, are found to be, respectively,
\begin{small}
\begin{equation}
\begin{aligned}
j^{\left(1\right)}_{y}&\!=\!\left[K_{yx}\partial_{x}T+\left(K_{yxx}
\cos\theta+K_{yxy}\sin\theta\right)B\right]\partial_{x}T\\
&=0,\\
j^{\left(2\right)}_{y}&\!=\!\left[W_{yxx}\!+\!\left(Q_{yxxx}\cos\theta\!+\!Q_{yxxy}\sin\theta\right)B\right]
\left(\partial_{x}T\right)^{2}\\
&=Q_{yxxx}\cos \theta \left(\partial_{x}T\right)^{2}B.
\end{aligned}
\label{current}
\end{equation}
\end{small}
To obtain Eq. (\ref{current}), we have used the equations in Eq. (\ref{coefficients}). Taking $a=y, b=c=d=x$ into Eq.(\ref{App-B-cu-f2}) and, meanwhile, using the relation in Eq. (\ref{App-B-Rel}), the quantity $Q_{yxxx}$ can be determined and is given in Eq.(\ref{Q_yxxx}). And the coefficients ( $\Upsilon_{1}$,$\Upsilon_{2}$,$\Upsilon_{3}$ and $\Upsilon_{4}$ ) in Eq.(\ref{Q_yxxx}) are found to be
\begin{equation}
\begin{aligned}
\Upsilon_{1}&=v^{2}_{x}v^{2}_{y},\\
\Upsilon_{2}&=2v_{y}v_{x}v_{xy}+2v_{xx}v_{y}^{2}+v^{2}_{x}v_{yy},\\
\Upsilon_{3}&=v_{xx}v_{yy}+v_{xxy}v_{y},\\
\Upsilon_{4}&=\left(v_{x}^{2}v_{yy}+2v_{x}v_{y}v_{xy}+v^{2}_{y}v_{xx}\right),
\label{App-B-Ga}
\end{aligned}
\end{equation}
$v_{a}=\partial \epsilon^{0}_\mathbf{k}/(\hbar \partial k_{a})$ denotes the $a$ component of electron velocity in absence of a magnetic field. Here, for simplicity, the coefficients $v_{ab}={\partial v_{a}}/{\partial k_{b}}$ and $v_{abc}={\partial v_{a}}/{\partial k_{b}\partial k_{c}}$ have been introduced. In the polar coordinate system ($k$,$\phi_{k}$), where $\phi_{k}$ is the polar angle measured from $k_{x}$ axis, one can obtain
\begin{equation}
\begin{aligned}
\epsilon^{0}_{k}=\sqrt{\epsilon_{1}^{2}+\eta^{2}\epsilon_{1}^{6}\cos^{2}3\phi_{k}},
\end{aligned}
\end{equation}
where $\epsilon_{1}=v_{F}\hbar k$ and $\eta=\lambda/(v_{F}\hbar)^{3}$. For $v_{x}$
\begin{equation}
\begin{aligned}
v_{x}&=\frac{v_{F} \epsilon_{1}\left[\cos\phi_{k}+1.5\eta^{2}\epsilon_{1}^{4}\left(\cos\phi_{k}+\cos5\phi_{k}\right)\right]}
{\epsilon^{0}_{k}},
\end{aligned}
\end{equation}
For $v_{y}$
\begin{equation}
\begin{aligned}
v_{y}&=\frac{v_{F} \epsilon_{1}\left[\sin\phi_{k}+1.5\eta^{2}\epsilon_{1}^{4}\left(\sin\phi_{k}-\sin5\phi_{k}\right)\right]}
{\epsilon^{0}_{k}},
\end{aligned}
\end{equation}
For $v_{xx}$, $v_{xy}$, $v_{yy}$ and $v_{xxy}$
\begin{small}
\begin{equation}
\begin{aligned}
v_{xx}=&\xi_{k}
\left[6\zeta^{2}\cos^{4}\phi_{k}\left(2\cos2\phi_{k}-1\right)^{3}+\sin^{2}\phi_{k}\right.\\
&\left.+\zeta\left(3.5+1.5\cos2\phi_{k}+
6\cos4\phi_{k}-\cos6\phi_{k}\right)\right],\\
v_{xy}=&\xi_{k}
\left[-6\zeta^{2}\cos3\phi_{k}\sin\phi_{k}-\frac{1}{2}\sin2\phi_{k}\right.\\
&\left.+1.5\left(\sin2\phi_{k}-4\sin4\phi\right)\zeta\right],\\
v_{yy}=&\xi_{k}\cos^{2}\phi_{k}
\left[-6\zeta^{2}\cos^{2}\phi_{k}\left(2\cos2\phi_{k}-1\right)^{3}\right.\\
&\left.+1+\zeta\left(15-16\cos2\phi_{k}-4\cos4\phi_{k}\right)\right],\\
v_{xxy}=&\frac{\epsilon_{1}\xi_{k}}{\!\left(\epsilon^{0}_{k}\!\right)^{2}}\!\left[\!48\zeta^{2}\cos^{6}
\phi_{k}\!\sin\phi_{k}\left(\cos^{4}\phi_{k}\!-\!9\sin^{4}\phi_{k}\right)\right.\\
&\left.+\frac{\zeta}{8}\left(35\sin\phi_{k}-162\sin3\phi_{k}+26\sin5\phi_{k}\right.\right.\\
&\left.\left.+7\sin7\phi_{k}\right)+\frac{1}{4}\left(3\sin3\phi_{k}-\sin\phi_{k}\right)\right].
\end{aligned}
\label{App-B-gamma}
\end{equation}
\end{small}
with $\xi_{k}=v_{F}^{2}\hbar\epsilon_{1}^{2}/{\left(\epsilon^{0}_{k}\right)^{3}}$ and $\zeta=\eta^{2}\epsilon_{1}^{4}$.
\setcounter{equation}{0}
\setcounter{figure}{0}
\setcounter{table}{0}
\makeatletter
\renewcommand{\theequation}{c\arabic{equation}}
\renewcommand{\thefigure}{c\arabic{figure}}
\renewcommand{\thetable}{c\arabic{table}}

\bigskip
\bigskip

\noindent
\section{The conversion rate from nonlinear spin to charge current }
\begin{figure}
\centering
\includegraphics[width=1.0\linewidth,clip]{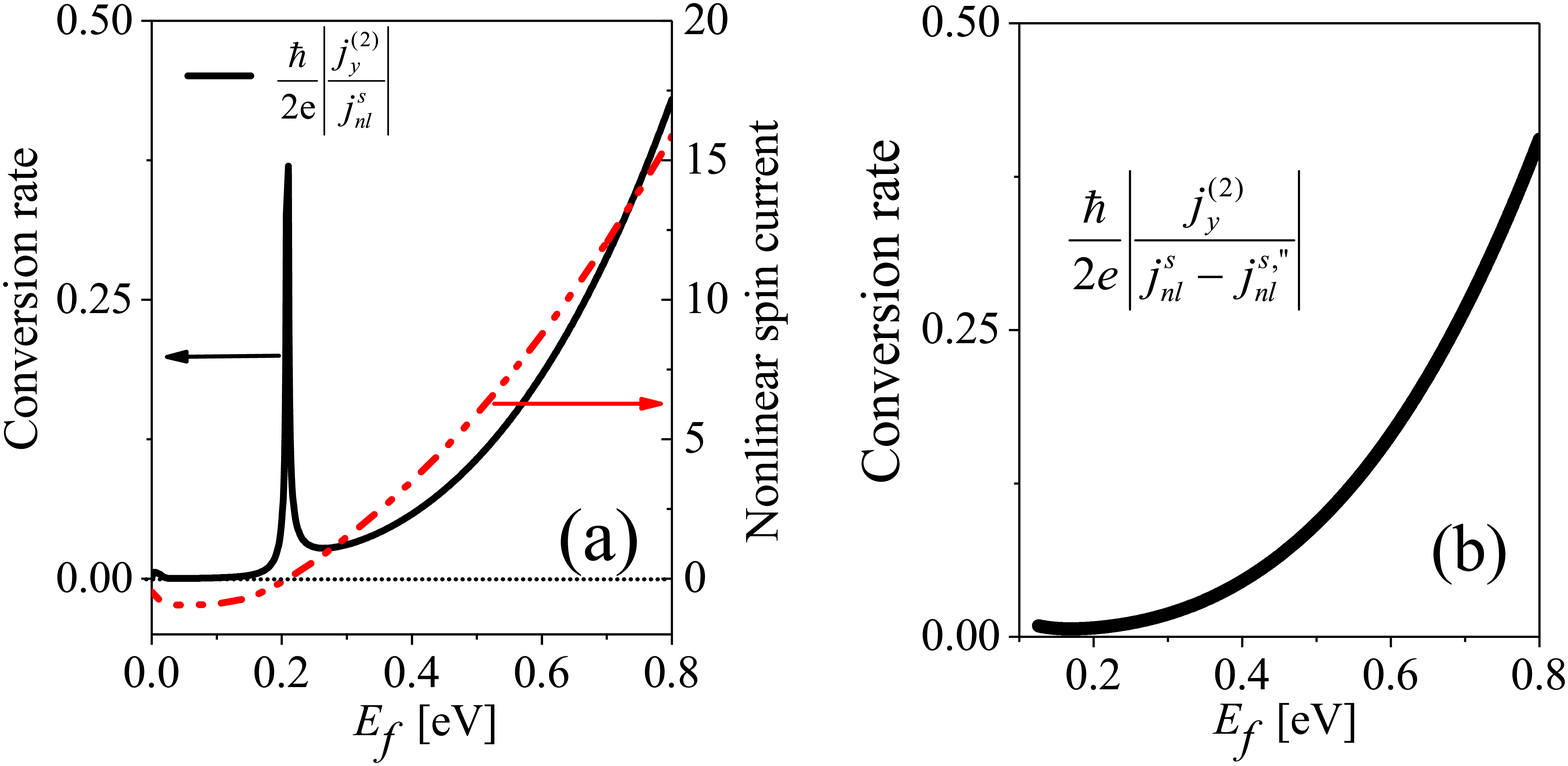}
\caption{(a) The conversion rate [black solid-line]  of nonlinear spin current (NSC) $j^{s}_{nl}$  to nonlinear planar Nernst current $j^{\left(2\right)}_{y}$ and $j^{s}_{nl}$ [ red dash-dot-dot line] against the Fermi energy $E_{f}$. (b) The conversion of $j^{s}_{nl}-j^{s,\prime\prime}_{nl}$ to nonlinear planar Nernst current vs Fermi energy $E_{f}$. $j^{s}_{nl}$ involves both contributions from hexagonal warping effect and linear Dirac dispersion. $j^{s,\prime\prime}_{nl}$ represents the NSC stemmed from the linear-$k$ Dirac dispersion, namely the nonlinear spin current for $\lambda=0$. Thus, $j^{s}_{nl}-j^{s,\prime\prime}_{nl}$ denotes the hexagonal-warping-effect-induced NSC.}
\label{CR}
\end{figure}

Figure \ref{CR} illustrates the conversion rate from the nonlinear spin to charge current. One might observe that an unexpected peak appeared around $E_{f}=0.21 eV$ [Black line in Fig. \ref{CR}(a)], in the regime where the nonlinear spin current $j^{s}_{nl}$ is almost zero and the warping effect is inapparent. The appearance of peak feature is actually reasonable since the nonlinear spin current  $j^{s}_{nl}=j^{s,\prime}_{nl}+j^{s,\prime\prime}_{nl}$  can be regarded as the sum of hexagonal-warping-induced nonlinear spin current $j^{s,\prime}_{nl}$ and $j^{s,\prime\prime}_{nl}$ from the linear-$k$ Dirac dispersion. The non-zero $j^{s,\prime\prime}_{nl}$ cannot be converted into the nonlinear charge current around $E_{f}=0.21 eV$ but the non-zero $j^{s,\prime}_{nl}$ could give rise to a faint signal of charge current, leading to a peak feature of conversion rate for $|j^{\left(2\right)}_{y}/j^{s}_{nl}|$. When subtracting the $j^{s,\prime\prime}_{nl}$ from the nonlinear spin current $j^{s}_{nl}$, the peak feature disappears and the conversion rate increases monotonously with the increase of Fermi energy [Fig. \ref{F-NSEC}(b)], as expected, since the hexagonal warping effect is enhanced with increasing the Fermi energy.

\setcounter{equation}{0}
\setcounter{figure}{0}
\setcounter{table}{0}
\makeatletter
\renewcommand{\theequation}{D\arabic{equation}}
\renewcommand{\thefigure}{D\arabic{figure}}
\renewcommand{\thetable}{D\arabic{table}}

\bigskip
\bigskip

\noindent

\section{The other possible contributions to NPNE} \label{ndf}
\subsection{The contribution of linear dispersion near the Dirac point with a few $k_{B}T$}\label{CLDN}
\begin{figure}
\centering
 \centering
 \includegraphics[width=0.45\textwidth]{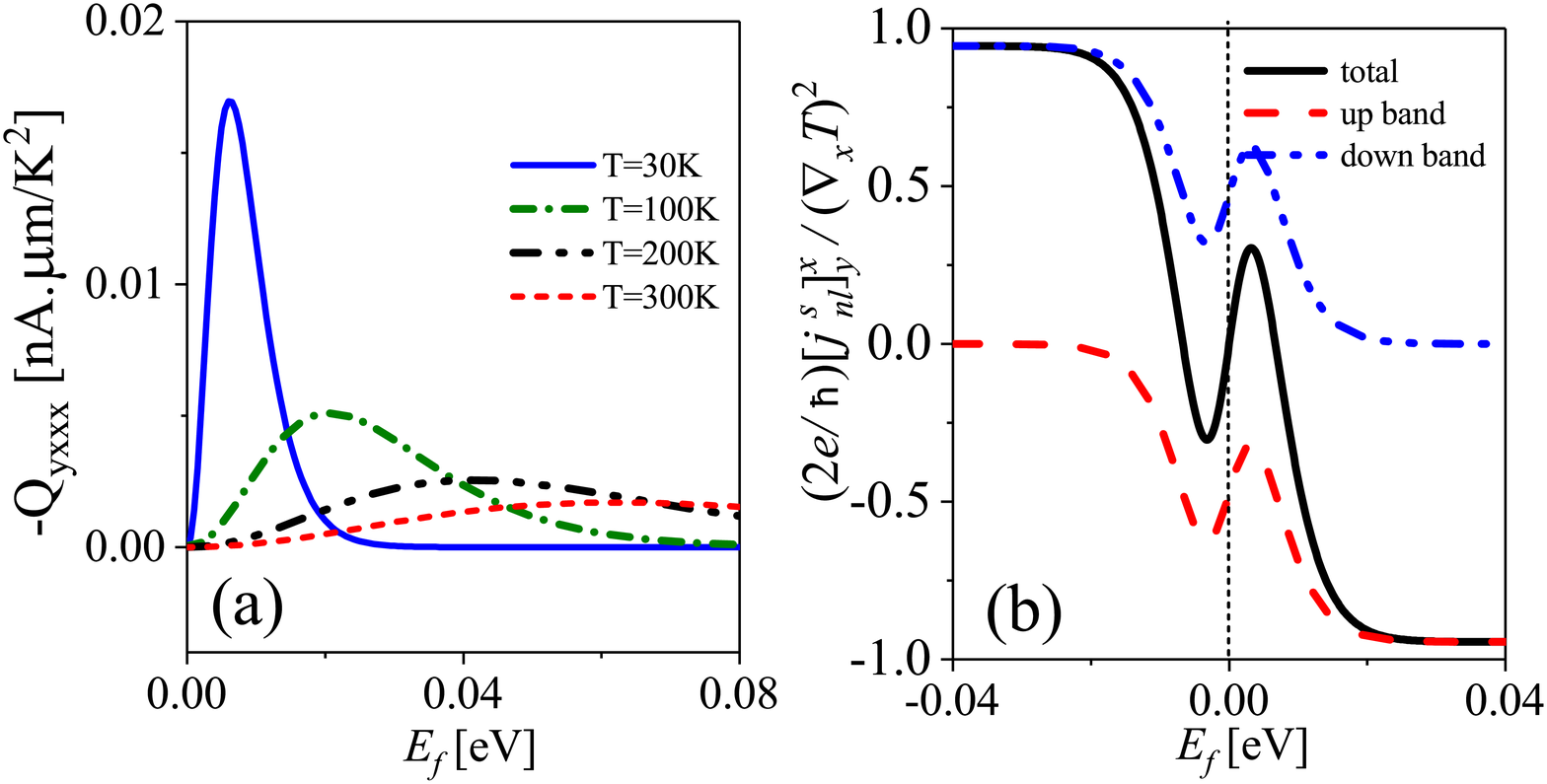}
 \centering
 \includegraphics[width=0.45\textwidth]{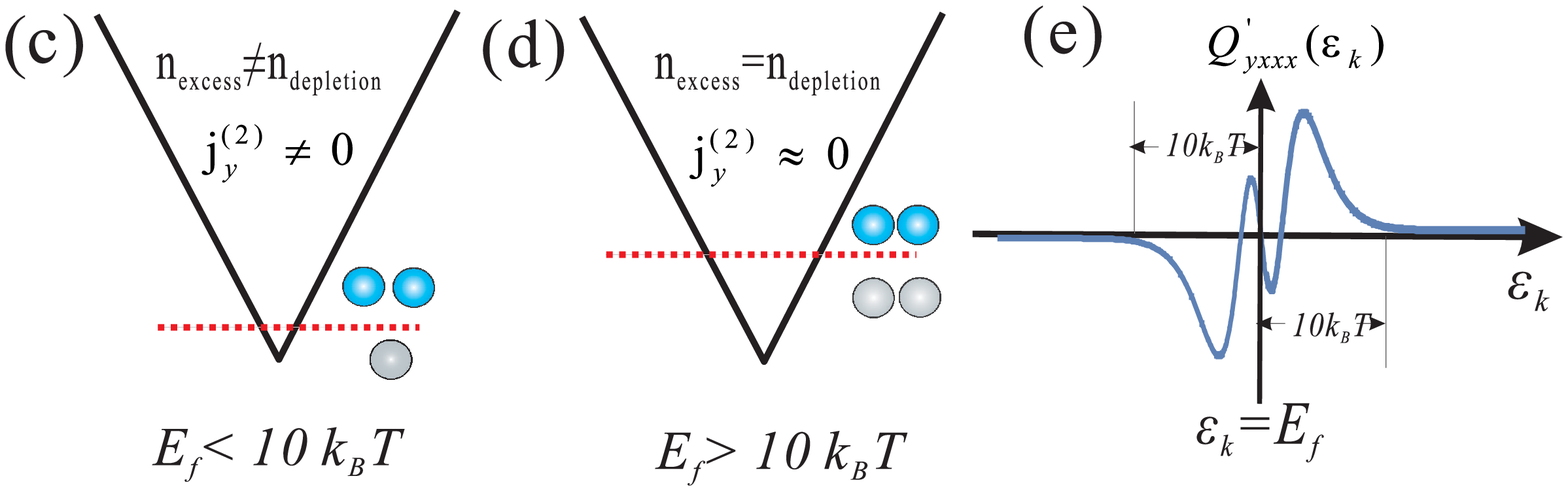}
\caption{(a) The nonlinear Planar coefficient (NPC) $Q_{yxxx}$ from the linear-$k$ Dirac dispersion as a function of Fermi energy for different temperature. (b)The nonlinear spin Nernst current $[{j}^{s}_{nl}]^{x}_{y}$ against Fermi energy $E_{f}$.  (c) (d)Schematic depiction for the generation of nonlinear planar Nernst effect from linear-$k$ Dirac dispersion. (e) $Q^{\prime}_{yxxx}(\epsilon_{k})$ vs $\epsilon_{k}$.}
\label{F-C-1}
\end{figure}
In this section, the faint signal [Fig. \ref{F-C-1}] arising from the linear dispersion near the Dirac point within a few $k_{B}T$ will be analysed. Letting the involved hexagonal warping term to be zero (i.e., $\lambda=0$) in quantities [ $\Upsilon_{1}$,$\Upsilon_{2}$,$\Upsilon_{3}$ and $\Upsilon_{4}$ ] and combining with a tedious derivation, the nonlinear planar coefficient $Q_{yxxx}$ [Eq.\ref{Q_yxxx}] originated from the linear dispersion can be determined as
\begin{equation}
Q_{yxxx}=\frac{\pi e\tau^{2}g\mu_{B}v_{F}}{4T^{2}\hbar^{2}}\int d\epsilon_{k} \left[Q_{1}\left(\epsilon_{k}\right)+E_{f}Q_{2}\left(\epsilon_{k}\right)\right],
\label{APP-D-Qyxxx}
\end{equation}
where
\begin{small}
\begin{equation}
\begin{aligned}
Q_{1}\left(\epsilon_{k}\right)&=(\epsilon_{k}-E_{f})\left(7\frac{\partial f_{0}}{\partial \epsilon_{k} }+Q_{2}\left(\epsilon_{k}\right)\right)+4(\epsilon_{k}-Ef)^{2}\frac{\partial^{2} f}{\partial \epsilon_{k}^{2}},\\
Q_{2}\left(\epsilon_{k}\right)&=\frac{\partial f_{0}}{\partial \epsilon_{k}}+3(\epsilon_{k}-E_{f})\frac{\partial^{2} f_{0}}{\partial \epsilon^{2}_{k}}+(\epsilon_{k}-E_{f})^{2}\frac{\partial^{3} f_{0}}{\partial \epsilon^{3}_{k}}.
\end{aligned}
\end{equation}
\end{small}
The quantities $Q_{1}\left(\epsilon_{k}\right)$ and $Q_{2}\left(\epsilon_{k}\right)$ are essentially zero when the energy  is beyond the range of $[E_{f}-10k_{B}T,E_{f}+10k_{B}T]$. When the Fermi level is larger than $10 k_{B}T$ [Fig. \ref{F-C-1} (d)], the term $Q_{1}\left(\epsilon_{k}\right)$ will have no contribution to the nonlinear planar Nernst effect owing to the anti-symmetry property, namely $Q_{1}\left(\epsilon_{k}+E_{f}\right)=-Q_{1}\left(\epsilon_{k}-E_{f}\right)$ [Fig.\ref{F-C-1} (d)]. For  $Q^{\prime \prime}\left(\epsilon_{k}+E_f\right)$ term, although it is an even function of $\epsilon_{k}$, it satisfies
\begin{equation}
\int_{E_{f}}^{E_{f}+10k_{B}T}d\epsilon_{k} Q_{2}\left(\epsilon_{k}\right)=\int_{E_{f}-10k_{B}T}^{E_{f}}d\epsilon_{k} Q_{2}\left(\epsilon_{k}\right)\approx 0
\end{equation}
Thus, when Fermi energy is larger than $10 k_{B}T$, $Q_{2}\left(\epsilon_{k}\right)$ also has no contribution to $Q_{yxxx}$. This is consistent with the result in the main text that the signal of nonlinear spin current originated from the linear-$k$ Dirac dispersion will not be converted into the nonlinear planar Nernst current.

Next, let us analyse the appearance of the weak signal near the Dirac point within a few $k_{B}T$, namely $E_{f}<10 k_{B}T$. In this regime, the contribution of  $E_{f} Q_{2}\left(\epsilon_{k}\right)$ term in Eq. (\ref{APP-D-Qyxxx}) to nonlinear planar Nernst coefficient $Q_{yxxx}$ can be neglected since $E_{f}$ can be viewed as a small quantity ($k_{B}T\approx 2.5 meV$ for $T=30 K$). The contribution to the nonlinear planar Nernst effect mainly come from $Q^{\prime}\left(\epsilon_{k}\right)$ . Figure \ref{F-C-1}(a) shows the variation of $Q^{\prime}\left(\epsilon_{k}\right)$ towards energy $\epsilon_{k}$. When the Fermi energy is located in the range of $[0, 10k_B T]$, there is no states in the range of $[E_{f}-10k_{B}T, 0 eV]$ for upper band. Therefore, the depleted or excessive carriers below the Fermi energy due to the second-order variation of temperature gradient and magnetic field are no longer equal to the excessive or depleted carriers above the Fermi energy. As a result, the carries are no longer in balance and lead to a weak signal of the nonlinear planar Nernst coefficient. Thus, the appearance of the weak signal from the linear-k Dirac dispersion could, physically, be attributed to the temperature broadening effect of nonequilbirum Fermi distribution near the Dirac point.




\subsection{Contribution of the particle-hole asymmetry}\label{APP-C-PHA}
In this section, we will discuss the contribution of the particle-hole asymmetry, namely $E_{0}(k)=\frac{\hbar^{2}k^{2}}{2m^{*}}$ term, to the nonlinear planar Nernst effect (NPNE). Unlike the contribution of particle-hole asymmetry to the nonlinear planar Hall effect (NPHE) in which the contributions related to particle-hole asymmetry and hexagonal warping are the same order of magnitude, we shall show below that the independent contribution of the particle-hole asymmetry to NPNE is insignificant. Explicitly, we start with the following model Hamiltonian without hexagonal warping effect for topological insulator in the presence of the in-plane magnetic field $\mathbf{H}$

\begin{figure}
\centering
\includegraphics[width=1.0\linewidth,clip]{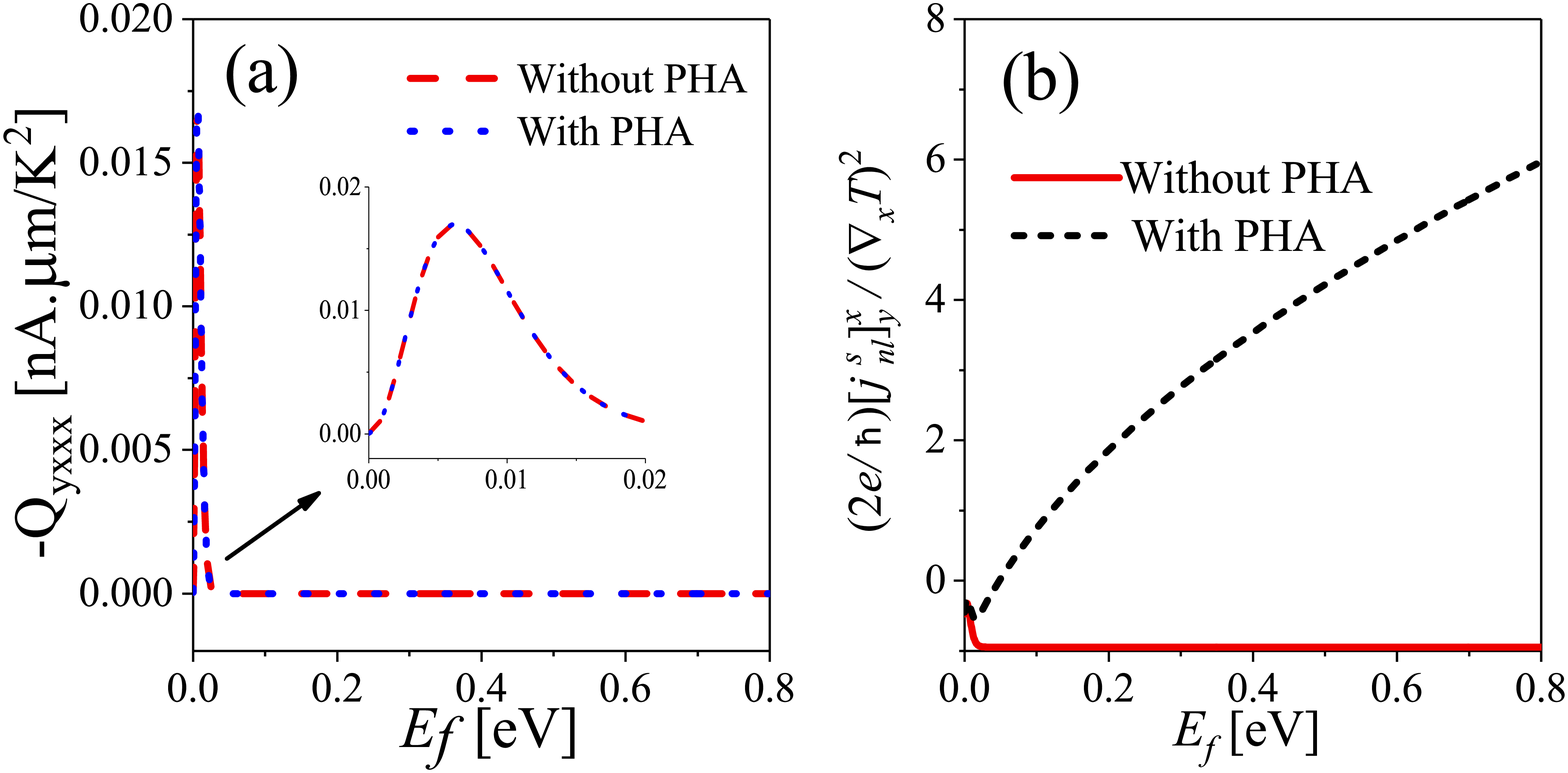}
\caption{The nonlinear planar coefficient $Q_{yxxx}$ [(a)] and the nonlinear spin Nernst current $[{j}^{s}_{nl}]^{x}_{y}$ [(b)] of upper band as a function of Fermi energy with or without particle-hole asymmetry (PHA). Inset: $Q_{yxxx}$ vs $E_{f}$ near the Dirac point. Parameters are used: $T=30 \text{K}$,  $v_{F}\hbar=2.25 \text{eV}{\AA}$, $m^{*}=0.09m$, $g=2$  and $\tau=5.864\times10^{-13}s $. Here, all parameters are taken from topological insulator Bi$_2$Te$_3$. }
\label{App-c-fi-ho}
\end{figure}
\begin{equation}
H^{\prime}=\frac{\hbar^{2}k^{2}}{2m^{*}}+v_{F}\hbar \boldsymbol{\sigma}\cdot(\mathbf{k}\times \hat{\mathbf{z}})+g u_{B}\boldsymbol{\sigma}\cdot\mathbf{H},
\label{APP-C-Ha1}
\end{equation}
The energy eigenvalues are
\begin{small}
\begin{equation}
\epsilon^{\text{M}}_{n}=\frac{\hbar^{2}k^{2}}{2m^{*}}+n\sqrt{(v_{F}\hbar k)^{2}+2v_{F}\hbar g u_{B}\mathbf{H}\cdot(\mathbf{k}\times {\hat{z}})+(gu_{B}H)^{2}},
\end{equation}
\end{small}
and the corresponding energy $\epsilon^{0}_{k}$ in Eq.(\ref{Q_yxxx}) without the perturbation of magnetic field for upper band is
\begin{equation}
\epsilon^{0}_{k}=\frac{\eta_{2}}{2}\epsilon_{1}^{2}+\epsilon_{1},
\end{equation}
where $\eta_{2}=\left(\frac{\hbar^{2}}{m^{*}}\right)/\left(v_{F}\hbar\right)^{2}$ and $\epsilon_{1}=v_{F}\hbar k$.
Thus, the corresponding quantities [$v_{x},\, v_{y}$,$\, v_{xx},\, v_{yy},\, v_{xy}\, v_{xxy}$] in Eq. (\ref{App-B-Ga}) are found to be
\begin{equation}
\begin{aligned}
v_{x}&=v_{F}\cos\phi_{k}\left(1+\eta_{2} \epsilon_{1}\right),\\
v_{y}&=v_{F}\sin\phi_{k}\left(1+\eta_{2}\epsilon_{1}\right),\\
v_{xx}&=v_{F}^{2}\hbar\left(\frac{\sin^{2}\phi_{k}}{\epsilon_{1}}+\eta_{2}\right),\\
v_{yy}&=v_{F}^{2}\hbar\left(\frac{\cos^{2}\phi_{k}}{\epsilon_{1}}+\eta_{2}\right),\\
v_{xy}&=v_{F}^{2}\hbar\left(-\frac{\sin2\phi}{2\epsilon_{1}}+\eta_{2}\right),\\
v_{xxy}&=\frac{v_{F}^{2}\hbar}{4\epsilon_{1}^{2}}\left(3\sin 3\phi_{k}-\sin \phi_{k}\right).
\end{aligned}
\label{APP-C-B-S}
\end{equation}
With Eqs (\ref{App-B-Ga})(\ref{APP-C-B-S}), the nonlinear planar Nernst coefficient (NPNC) $Q_{yxxx}$  [Eq. (\ref{Q_yxxx})] quantizing the NPNE can be determined. Figure \ref{App-c-fi-ho} (a) shows that the variation of $Q_{yxxx}$ as a function of Fermi energy with or without particle-hole asymmetry (PHA) ${\hbar^{2}k^{2}}/{2m^{*}}$. It is found that with or without the particle-hole asymmetry makes no difference to the magnitude of $Q_{yxxx}$ [Fig. \ref{App-c-fi-ho} (a)] in Bi$_{2}$Te$_{3} $ , which means that the particle-hole asymmetry can not independently give rise to NPNE that is distinguish from the nonlinear planar Hall effect. In fact, the weak signal appeared near the Dirac point within 25 meV ($\sim 10 k_{B}T$ for $T=30 K$) is originated from the linear-$k$ Dirac dispersion and induced by the thermal broadening effect [see detail in Sec. \ref{CLDN} ]. Parameters are used for Bi$_{2}$Te$_{3}$: the Fermi velocity $v_{F}\hbar=2.25\,\text{eV} {\AA}$ and $m^{*}=0.09 m_{e}$ \cite{Zhang,Liu} where $m_{e}$ is free electron mass.
 \begin{figure}[h]
 \centering
 \includegraphics[width=0.48\textwidth]{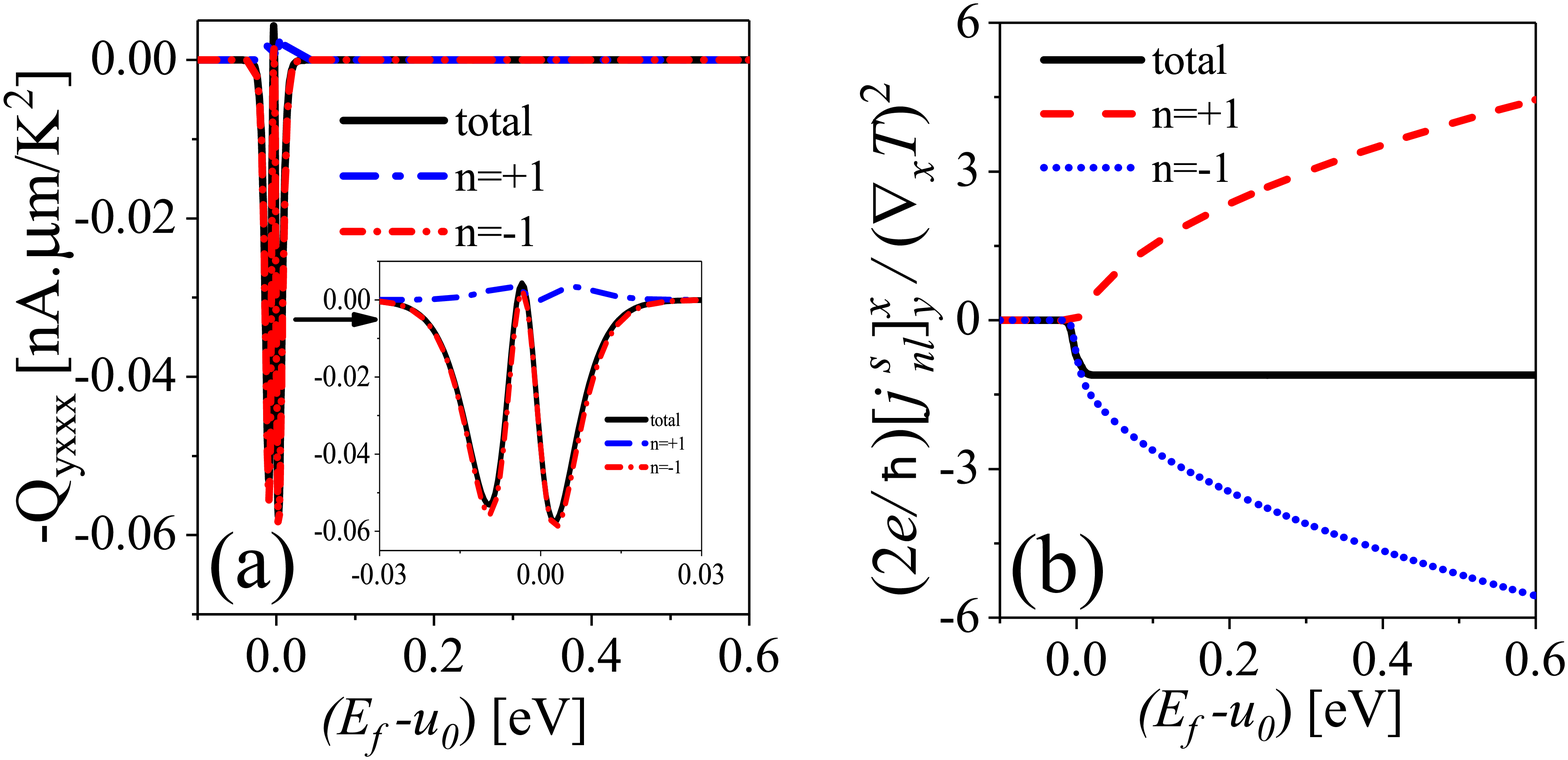}
  \begin{minipage}[t]{0.48\linewidth}
 \centering
 \includegraphics[width=1.1\textwidth]{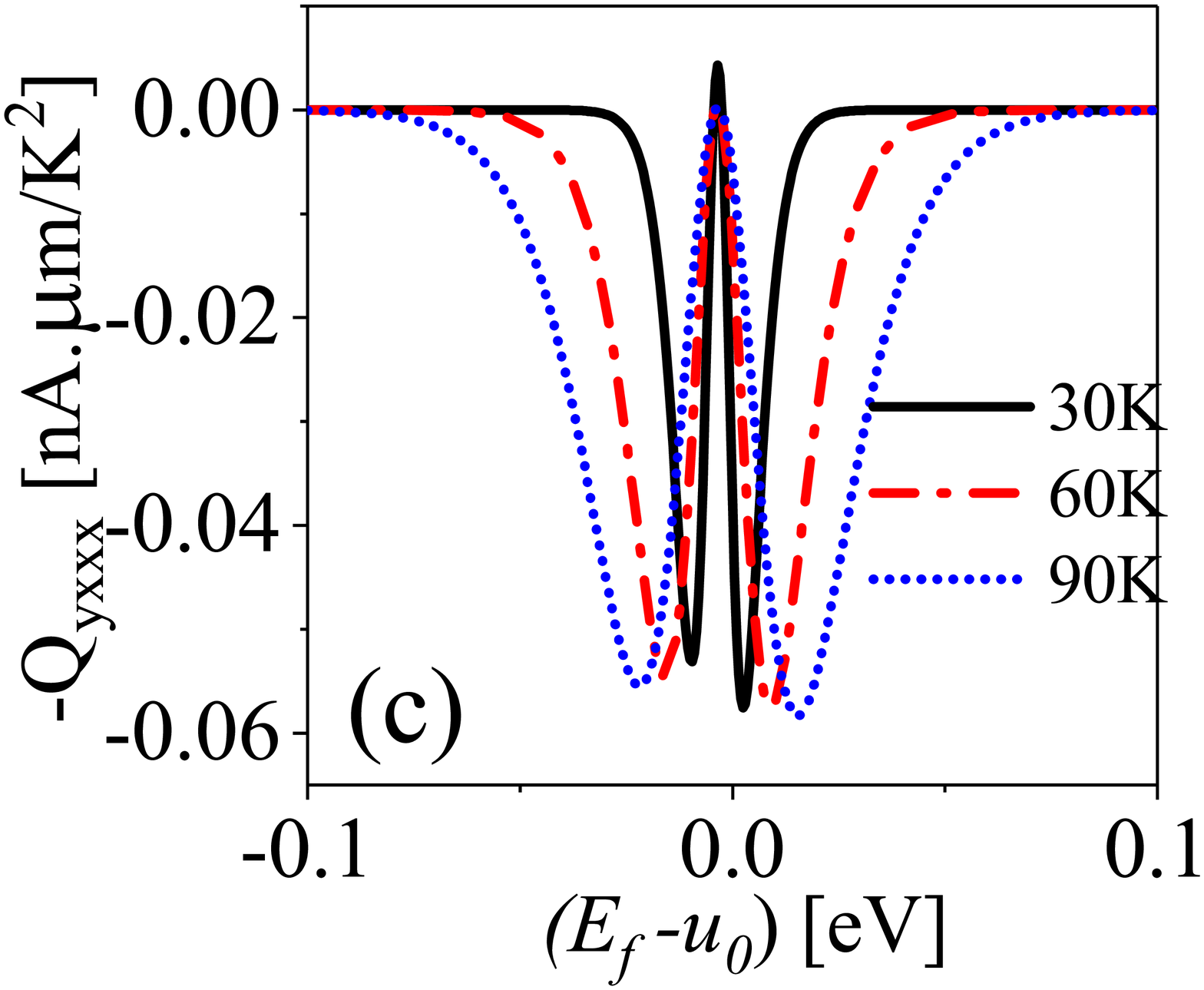}
 \end{minipage}
 \hfill
 \begin{minipage}[t]{0.45\linewidth}
 \centering
 \includegraphics[width=0.85\textwidth]{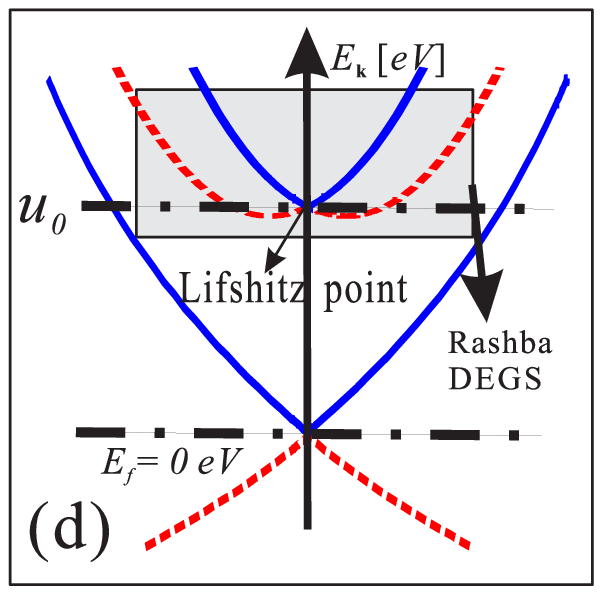}
 \end{minipage}

\caption{The nonlinear planar coefficient $Q_{yxxx}$ [(a)] and the nonlinear spin Nernst current $[{j}^{s}_{nl}]^{x}_{y}$ [(b)] from the Rashba-split surface states in two-dimensional electron gas (2DEG) as a function of Fermi energy.  The red dashed line and the blue short-dot line are the contributions from upper $n=+1$ and down $n=-1$ subbands, respectively. The black solid-line is the sum of the two contributions. (c) $Q_{yxxx}$ vs $E_{f}$ for different temperature. (d) Schematic depiction of the band structures for the surface states of topological insulators, where the shaded part indicates the Rashba 2DEG due to the surface band bending. $T=30 K$ is fixed in (a) and (b). Parameters are used: $m^{*}_{e}=0.2 m_{e}$ with bare mass of an electron $m_{e}$, $\alpha_{R}=0.5 \text{eV}{\AA}$ and g=2.}
\label{App-c3}
\end{figure}
\subsection{Contribution of the Rashba 2DEG}\label{COR2D}
Due to the surface band bending in topological insulator, a Rashba-split surface states in two-dimensional electron gas (2DEG) [Fig. \ref{App-c3}(d)] might coexist with topological surfaces states (TSS). To theoretically investigate the contribution of a Rashba to nonlinear planar Nernst effect, we begin with the following model Hamiltonian

\begin{equation}
H^{\prime}=u_{0}+\frac{\hbar^{2}k^{2}}{2m^{*}_\text{R}}+\alpha_{R}\hbar \boldsymbol{\sigma}\cdot(\mathbf{k}\times \hat{\mathbf{z}})+g u_{B}\boldsymbol{\sigma}\cdot\mathbf{H},
\label{APP-C-Ha2}
\end{equation}

where $m^{*}_\text{R}$ represents the effective mass, $u_{0}$ is chemical potential, and  $\alpha_{R}$ denotes the strength of the Rashba spin-orbit coupling. The nonlinear planar Nernst coefficient $Q_{yxxx}$ [Eq.\ref{Q_yxxx})] with an in-plane magnetic field and spin Nernst current $[j^{s}_{nl}]^{x}_{y}$ without a magnetic field as the second-order response to temperature gradient can be obtained accordingly based on the Hamiltonian Eq. (\ref{APP-C-Ha2}) in the same manner as Sec.\ref{APP-C-PHA}.

Figure \ref{App-c3} (a) shows the contributions of the $n=+1$ and $n=-1$ subbands (namely, the inner and outer Fermi contours of Rashba 2DEG) to nonlinear planar Nernst effect. It is found that only when the Fermi energy located near the Lifshitz point [Fig. \ref{App-c3} (d)] within a few $k_{B} T$ [Fig. \ref{App-c3} (a)(c)], a nonzero $Q_{yxxx}$ can be generated and is almost $100$ times smaller than the signal from TSS. In fact, the appearance of this faint signal might be attributed to the temperature broadening effect like the signal stemmed from the linear-k Dirac dispersion [see details in Sec. \ref{CLDN}]. When modulating  Fermi energy away from the Lifshitz point, the nonlinear planar Nernst effect disappears since there is no nonlinear spin current converted into charge current for both subbands.

\end{document}